\documentstyle[11pt,newpasp,twoside]{article}

\def\cge      {{$_ >\atop{^\sim}$}}
\def\cle      {{$_ <\atop{^\sim}$}}

\def\square {\hbox{\vrule width5pt height5pt}}
\def\itm#1 {\vskip10pt \noindent \square\ {\bf #1} }
\def\arcm     {{\ifmmode {'\ }\else$'     $\fi} } 
\def\arcs     {{\ifmmode{''\ }\else$''    $\fi} } 

\def\Lsun     {{$L_{\odot}$} }

\def\Msun     {{\ $M_{\odot}$} }

\def\Msun     {{\ $M_{\odot}$} }

\def\edcomment#1{\iffalse\marginpar{\raggedright\sl#1\/}\else\relax\fi}
\reversemarginpar

\begin{document}

\title{High Resolution Radio Imaging of Distant Submillimeter Galaxies}

\author{E. A. Richards} 


\begin{abstract}
        Using a combination of radio and optical imaging
at 0.1\arcs to 0.2\arcs resolution with the VLA/MERLIN and HST
has led to a breakthrough in our understanding of radio emission
from distant (0.1$<z<$3) starburst galaxies. We have recently isolated
a  number of high redshift, dusty
starburst galaxies that remain invisible in
ground based images to $I_{AB}$=25
and $I_{AB}$=28.5 in the Hubble Deep Field.
These galaxies appear as faint radio sources,
often accompanied by very red counterparts
($I-K > 4-6$) and  submillimeter sources
with S > 2 mJy at 850 microns as measured
with SCUBA on the JCMT.
The far-infrared luminosities of these galaxies exceeds
even the most intense starbursts found in
the local universe (e.g., Arp 220), suggesting
they are in the process of converting the
bulk of their gass mass into stars.                  

        These galaxies,
completely absent in optical surveys,
constitute 50\% - 90\% of the star-formation
density in the distant Universe. Given the
poor sub-mm resolution (15\arcs ) of the
SCUBA/JCMT images, we use the
0.2\arcs radio imaging as a surrogate
in order to understand the nature of the
dominant emission mechanism driving the
FIR luminosity (AGN vs. star-formation).
Upcoming developments in radio instrumentation
(the Expanded VLA Array and Square Kilometer
Array) will soon increase sensitivity and
resolution orders of magnitude, providing a
natural complement  to parallel developments in
sub-mm facilities (e.g., ALMA). With dual
radio continuum and sub-mm surveys of the
distant Universe, a census of galaxy evolution
to the earliest cosmic epochs (z = 5-30)
will soon be possible.                         

\end{abstract}

\vspace*{-1mm}
\section*{Radio Emission from Distant Galaxies in the HDF }
\vspace*{-1mm}                                   

         The diffuse radio emission observed in local
starbursts is believed to be a mixture of
synchrotron radiation (excited by supernovae
remnants and hence directly proportional
to the number of supernovae producing stars)
and thermal radiation (from HII regions and hence
an indicator of the number of O and B stars in a galaxy).
As the thermal and synchrotron radiation of a
starburst dissipates on a
physical time scale of $10^7-10^8$ years, the
radio luminosity is a true measure of the
instantaneous star-formation rate (SFR) in a galaxy, uncontaminated by
older stellar populations.  Since supernovae
progenitors are dominated by $\sim$8 \Msun stars,
synchrotron radiation has the additional advantage of being
less sensitive to uncertainties in the initial
mass function as opposed to UV and optical           
recombination line emission. However, the most
obvious advantage of using the radio luminosity
as a SFR tracer is its unsusceptibility to dust
obscuration, as galaxies and the inter-galactic
medium are transparent at centimeter wavelengths.
The strong correlation between far-infrared and
radio emission from local star-forming galaxies
suggests that radio emission from distant, dust
obscured galaxies should be visible at the microjansky
level for luminous starbursts at redshifts less than
3-4.

        We have recently completed a deep radio survey of
the Hubble Deep Field using both the Multi-Element Microwave Linked
Interferometer (MERLIN) and the Very Large Array (VLA)
at 1.4 and 8.5 GHz (Richards et al. 1998, AJ , 116, 1039; 
Richards  1999a, ApJL, 511, 1; Richards 1999b, ApJ, 2000, in press,
astro-ph/9908313;
Muxlow et al. 2000, in prep.) in order to study the
nature of microjansky radio galaxies,
and in particular understand their implication for galaxy
evolution at early epochs. The optical identifications of the
72 radio
sources detected in a complete sample ($S_{1.4} \geq$ 40 $\mu$Jy or     
6$\sigma$)
on the HST images in the HDF and flanking fields show that:

\vspace{-4pt}
\itm{ 70$\pm$10\% of the optical identifications
  are associated with morphologically peculiar, merging and/or interacting
  galaxies, many with independent evidence for active star-formation
  (blue colors, infra-red excess, HII-like emission spectra).}

\vspace{-4pt}
\itm{ The remaining identifications are composed
 of low-luminosity FR Is, Seyferts, LINERs, and luminous star-forming
field spirals at low redshift (representative identifications are shown
in Figure 1).}

\vspace{-4pt}
\itm{ The radio spectral indices are in general
  steep ($\alpha > 0.5$ ; $S \propto \nu ^{-\alpha }$) and
  the median radio angular size about 1-1.5\arcs , indicative
  of diffuse synchrotron emission in $z = 0.2-1.3$ galactic disks.}
                                                             
\vspace{-4pt}
\itm{ 20\% of the radio sources cannot be
  identified to $I_{AB}$ = 25 in deep ground based
images and to $I_{AB}$ = 28.5 in the HDF itself. These radio
sources are likely distant, extreme starburst systems
enshrouded in dust. This 'new' population is discussed in
more detail in Richards et al. (1999, ApJ, in press, astro-ph/9909251)
and one source in particular, observed with HST-NICMOS by 
Waddington et al. (1999, ApJ, in press, astro-ph/9910069). 
Similar radio sources have recently been 
reported in the HDF-S (Norris et al. 1999, astro-ph/9910437).}

Thus the cosmological faint radio population
is dominated by the distant analogs
of local IRAS galaxies with suggested
star-formation rates of 10-1000 \Msun yr$^{-1}$. In principle
this radio selected starburst population allows for 
a derivation of the star-formation history, independent
of optical selection biases.

\vspace*{-1mm}
\section*{Detection of Distant Ultraluminous Radio Selected
Starburst Galaxies}
\vspace*{-1mm}      
           
   In March and June 1999, we 
obtained shallow JCMT/SCUBA images of
14 optically faint radio sources in the Hubble
Flanking fields. We detected 5 of these sources
above 6 mJy at 850 $\mu$m. None of the 32 lower redshift
(0.2 $< z <$ 1.3) radio sources in our field of view were
detected. Comparison of our source counts with those
from previous sub-mm surveys (Eales et al. 1999; Hughes et al.
1999; Barger, Cowie \& Sanders 1999), shows that
our radio selection technique recovers
essentially all of the bright ($S_{850}$ \cge
6 mJy) sub-mm sources. Thus there is an almost one-to-one
correspondence between the bright sub-mm sky and
the optically invisible microjansky radio sources
($S_{1.4}$ \cge 40 $\mu$Jy). 

        Based on the far-infrared
to radio flux relationship observed in local starburst galaxies
such as Arp 220, we modeled the
redshifts of these optically faint, radio/sub-mm
galaxies and found they likely lie at 1 \cle $z$ \cle 3
(Carilli \& Yun 1999, ApJL, 1999, 513, 13).
We can use the sub-mm flux alone to estimate the overall
luminosity, only weakly dependent on redshift because
of the offsetting effects of far-IR spectral index
and cosmological dimming.
These values imply we are detecting ultraluminous
infrared galaxies with 10$^{12-13}$\Lsun, substantially 
more luminous than Arp 220. In the volume probed by
our survey between 1 \cle $z$ \cle 3 , this corresponds to a
volume averaged star-formation
rate of 0.4 \Msun /yr/Mpc$^3$, equivalent to the
{\em dust corrected} optical value obtained by Steidel
et al. (1999, ApJ, 519, 1). Thus optically 'invisible' objects
form an important constituent of the $z > 1$ star-formation
history, as shown by Barger, Cowie and Richards (1999, AJ submitted).
These
high redshift radio selected
starburst galaxies are completely missing from the
optical samples, implying that optical surveys
give a biased view of the distant star-forming
Universe.

    Our low redshift (0.1 $< z <$1.3) sample contains important
 information on the star-formation history as well.
 Based on the optical identifications in our deepest
 radio surveys, we have attempted to make a first guess
 of the radio determined star-formation history.
We use both the radio properties, such as spectral index and morphology,
as well as the optical morphology provided by HST images to cull a
clean sample of star-forming systems (Richards et al. 1998;
Richards 1999b; Haarsma et al. 1999).  Our preliminary results
are in general agreement with the dust corrected estimates of
Steidel et al., although systematically higher at all redshifts.
We interpret this as evidence of missing star-formation
is the optical studies due to underestimates of the
dust extinction. We cannot completely rule out
the possibility that our radio samples have
some contamination by low-luminosity AGN (i.e.,
Seyferts) which could also bring the radio and
optical surveys into better agreement.            

Deeper high resolution radio observations, with complete
spectroscopic coverage are needed to reveal the
amount of 'hidden' star-formation in the distant Universe. 
Only be combining, optical, radio, and far-infrared/sub-mm
measurements of distant galaxies can a reliable consensus
of their star-forming properties be obtained.

{                                                    
\small
        It is a pleasure to thank my collaborators
F. Bauer, A. Barger, L. Cowie, 
K. Kellermann, E. Fomalont, B. Partridge,
R. Windhorst, T. Muxlow,  I. Waddington and D. Haarsma.
Support for part of
this work was provided by NASA through grant 
HF-01123.01-99A from the STSI, which is operated by AURA,
Inc., under NASA contract
NAS5-2655.}

\vskip5pt

{\bf Figure Caption 1:}

 Montage of radio/HST flanking field overlays.
Contours are 1.4 GHz fluxes drawn at 2, 4, 8, 16,
 32, 64 $\sigma$ ($\sigma$ = 4 $\mu$Jy). Greyscale
is log strectch of HST I-band image 5\arcs
on a side. {\bf Upper Left:} I = 19.5 elliptical
with weak radio AGN core at $z = 0.32$. Twenty
percent of the IDs are AGNs.
{\bf Upper Right:} I = 21.1 disk galaxy is at $z = 0.96$
with a flat radio spectral index ($\alpha$ = 0.2). 
The optical spectra shows broad high excitation lines,
suggesting the presence of a Seyfert core.
{\bf Lower Left:} A dramatic
$z = 0.5$ merger with a starburst core. This galaxy
has about 1/3 the luminosity of Arp 220. About
60\% of the radio IDs are of this variety. 
A rather bright I = 18.3 mag disk galaxy with an
unusually steep radio spectrum of $\alpha > $ 1.6.
There is no known star-forming, or spiral galaxy
in the local Universe with such a steep non-thermal
spectrum. 
About 10\% of the radio sources fit into
this ultrasteep class.

\end{document}